\documentclass[aps,prl,twocolumn,superscriptaddress]{revtex4-1}
\usepackage{graphicx}
\usepackage{amsmath}
\usepackage{hyperref}

\begin{document}

\title{Non-equilibrium scale invariance and shortcuts to adiabaticity in a one-dimensional Bose gas}

\author{W. Rohringer}
\author{D. Fischer}
\author{F. Steiner}
\affiliation{Vienna Center for Quantum Science and Technology, Atominstitut, TU Wien, 1020 Vienna, Austria}
\author{I. E. Mazets}
\affiliation{Vienna Center for Quantum Science and Technology, Atominstitut, TU Wien, 1020 Vienna, Austria}
\affiliation{Ioffe Physical-Technical Institute of the Russian Academy of Sciences, 194021 St. Petersburg, Russia}
\affiliation{Wolfgang Pauli Institute, 1090 Vienna, Austria}
\author{J. Schmiedmayer}
\author{M. Trupke}
\affiliation{Vienna Center for Quantum Science and Technology, Atominstitut, TU Wien, 1020 Vienna, Austria}

\date{\today}

\begin{abstract}
We present experimental evidence for scale invariant behaviour of the excitation spectrum in phase-fluctuating quasi-1d Bose gases after a rapid change of the external trapping potential. Probing density correlations in free expansion, we find that the temperature of an initial thermal state scales with the spatial extension of the cloud as predicted by a model based on adiabatic rescaling of initial eigenmodes with conserved quasiparticle occupation numbers. Based on this result, we demonstrate that shortcuts to adiabaticity for the rapid expansion or compression of the gas do not induce additional heating.
\end{abstract}


\maketitle

A systematic understanding of non-equilibrium dynamics in many-body quantum systems is a longstanding goal, with far-reaching applicability for many different fields of physics. Ultracold atom experiments offer clean implementations of systems that are tunable, well isolated from the environment and theoretically tractable \cite{Polkovnikov2011,Rigol2008}. In particular, the profound understanding available for the one-dimensional (1d) Bose gas makes it an ideal test bed for quantum many-body dynamics \cite{Cazalilla2011}. 

Tunable parameters in the system's Hamiltonian allow the controlled preparation of non-equilibrium states \cite{Kinoshita2006, Sadler2006, Cheneau2012, Gring2012}. The identification of characteristic scaling laws is an important step for the concise description of the subsequent dynamical processes. Of particular importance are laws governing not only global parameters \cite{Kagan1996, Castin1996, Chevy2002} but ideally the full spectrum of excitations, as studied in recent experiments with 2d Bose \cite{Pitaevskii1997, Hung2011} or Tonks-Girardeau gases \cite{Minguzzi2005, Kinoshita2006}. 

Recent work \cite{Gritsev2010} has shown that a general scaling property of many-body wavefunctions holds exactly for a broad class of systems, including the weakly interacting 1d Bose gas addressed in this Letter. The existence of such a scaling solution is a consequence of a dynamical symmetry of the underlying Hamiltonian.
For an ultracold gas, fast changes of control parameters in the Hamiltonian generally lead to quasiparticle production and heating \cite{Fedichev2004a}. The existence of a scaling solution for the full spectrum of quasiparticle modes implies that so-called shortcuts to adiabaticity (STA) \cite{Chen2010b, Schaff2011a} can be engineered not only for the mean density profile of a 1d gas, but also for correlation properties of the system in certain regimes of interaction strength \cite{DelCampo2011, DelCampo2013}.

 We show in this work that the scaling solutions for a true many-body wavefunction have their counterpart in the hydrodynamic regime of our experimental system.  We bring our system out of equilibrium by rapidly changing its longitudinal confinement. The subsequent system evolution gives insight into the scaling properties of the gas. This allows us to study the regimes and limits of such a manipulation, with an emphasis on STA schemes. We furthermore demonstrate for the first time that STA schemes are valid for the second-order correlation, and thereby the temperature, of weakly interacting 1d Bose gases.

\section{Results and Discussion}

 \begin{figure} %\hspace*{-10px}
 \includegraphics[width=0.45\textwidth]{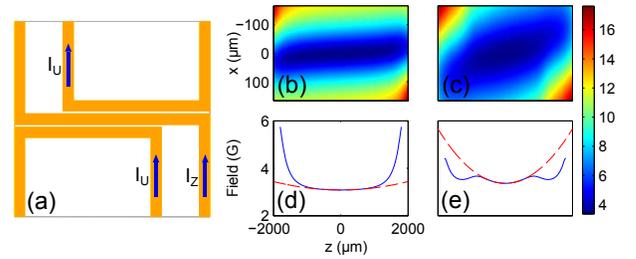}%
 \caption{Time-dependent potentials on an atom chip. (a) The current ratio between a central Z-shaped wire and two U-shaped control wires allows us to precisely tune the trap geometry. For a symmetric current flow, the trap minimum is positioned below the center of the Z-wire, with the long trap axis aligned to the horizontal direction. (b) 2d cut through the trapping potentials for $I_Z = 2 A, I_U = 0 A$ and (c) $I_Z = 1.5 A, I_U = 1 A$ at a constant external Bias field of $B = 26 G$, respectively. (d,e) Cuts through the radial trap minimum of the same potentials to show the axial trap deformation.}
 \label{fig1}
 \end{figure}

In our experiments, we investigate the scaling solutions of hydrodynamic equations and how they can be applied for the rapid control of the complete wavefunction of a many-body quantum system.

We start with a single quasicondensate of several thousand \textsuperscript{87}Rb atoms in an elongated trap on an atom chip \cite{Reichel2010}. The initial temperatures are set between 50 nK and 150 nK and linear densities range between 50 atoms/$\mu$m and 200 atoms/$\mu$m. Axially, the cloud is deeply in the Thomas-Fermi regime. Radially, the gas is described by an interaction-broadened ground state wavefunction \cite{Salasnich2002, Kruger2010, Amerongen2008}. For these parameters, both the chemical potential and the average thermal energy per particle fulfil the condition $\mu, k_BT \leq \hbar\omega_r$, where $\hbar\omega_r$ denotes the radial level spacing of the trap with frequency $\omega_r$, so that scattering into radial excited states is strongly suppressed and an effective 1d system is realized \cite{Goerlitz2001, Reichel2010, Stringari1998,Salasnich2002}. After evaporative cooling, we keep an RF-shield 12 kHz above the trap bottom throughout our experiments to remove hot atoms. The cloud is probed by standard absorption imaging techniques after a 4 ms to 10 ms long phase of time-of-flight expansion.

The geometry of the trap is governed by the current flow through a central Z-shaped wire and two U-shaped control structures on the atom chip, as shown in figure \ref{fig1}(a). Panels (b)-(e) show two different trapping potentials calculated for currents tuned to $I_Z = 2$ A and $I_U = 0$ A, as well as $I_Z = 1.5$ A and $I_U = 1$ A. Varying $I_Z$ and $I_U$ results in traps with axial confinement ranging from $\omega_a = 2\pi\times 16$ Hz to $\omega_a = 2\pi\times 7$ Hz, and radial confinement from $\omega_r = 2\pi\times 600$ Hz to $\omega_r = 2\pi\times 1100$ Hz.  A rapid change of the current ratio $I_U/I_Z$ constitutes a quench of the trapping potential and induces excitations. 

In our first set of experiments we probe the dynamical scaling of the phonon ensemble in the presence of an axial quadrupole-mode collective excitation \cite{Menotti2002} induced by such a quench. To this end, we employ a linear ramp from $\omega_a =$ $2\pi\times 12.1$ Hz to $2\pi\times 8.2$ Hz, and from $\omega_r = $ $2\pi\times 630$ Hz to $2\pi\times 990$ Hz, respectively, of duration $\tau$. The ramps of the trapping potential were designed to avoid transverse excitations. We chose to maintain a constant transverse position to avoid inducing a corresponding sloshing of the cloud. The ramp duration was chosen to be longer than $\tau \approx$ 5 ms so that adiabaticity with respect to the change of transverse trap frequency is fulfilled. Axial dipole oscillations are suppressed by the symmetric arrangement of the control wires.

\begin{figure}
 \includegraphics[width=0.48\textwidth]{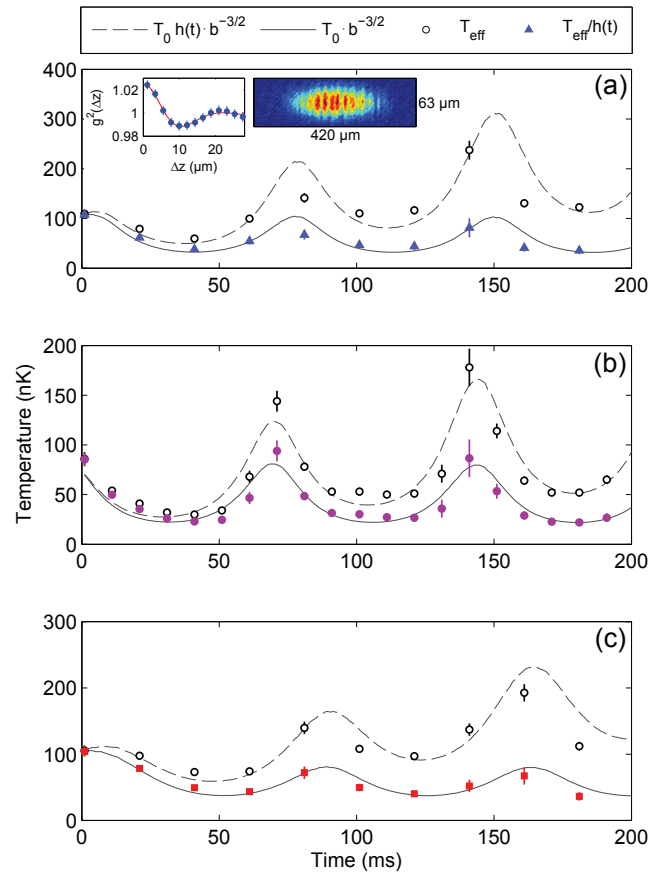}
 \caption{Temperature evolution following a quench. Black circles: temperatures measured from density correlations in free expansion. Dashed lines: scaling law taking into account heating as described by the expression $h(t)$ discussed in the methods section, fitted for effective rates for each dataset. Blue triangles, purple circles and red squares: temperatures corrected for heating rate. Lines: scaling law $T(t) = T(0)\cdot b^{-3/2}$ as discussed in the main text. Error bars represent the standard error estimated by a bootstrapping technique, as used in \cite{Kuhnert2013}. (a) Quench time $\tau =$ 10 ms, atom number N $\approx$ $16\cdot 10^3$ $\pm$ $10^3$, heating rate $\alpha\cdot T(0) \approx$ 0.54 nK/ms.  Inset: thermometry with density correlations in free expansion. Data points correspond to an average of autocorrelations over 350 density profiles integrated from pictures as depicted here. (b) $\tau =$ 10 ms, N $\approx 11\cdot 10^3$ $\pm$ $10^3$, heating rate $\alpha\cdot T(0) \approx$ 0.28 nK/ms. (c) $\tau =$ 30 ms, N $\approx$ $16\cdot 10^3$ $\pm$ $10^3$, $\alpha\cdot T(0) \approx$ 0.55 nK/ms.}
 \label{fig2}
 \end{figure}

We probe phononic excitations in the quasicondensate using a thermometry scheme based on the analysis of density correlations in free expansion \cite{Manz2010, Imambekov2009}, as shown in the inset of figure \ref{fig2}(a).  To extract the temperature we compare the measured density correlation functions with the results of a stochastic model \cite{Stimming2010}. Our analysis accounts for the effects of the collective excitation on the free expansion (see methods section below), and for the finite resolution of our imaging system. 

Figure \ref{fig2} summarises our temperature measurements following a quench. We show data for ramp times of 10 and 30 ms and mean atom numbers of 11000 and 16000, compared to the behaviour expected from a scaling model building upon the results of Ref. \cite{Gritsev2010}.

 \begin{figure} %\hspace*{-10px}
    \includegraphics[width=0.48\textwidth]{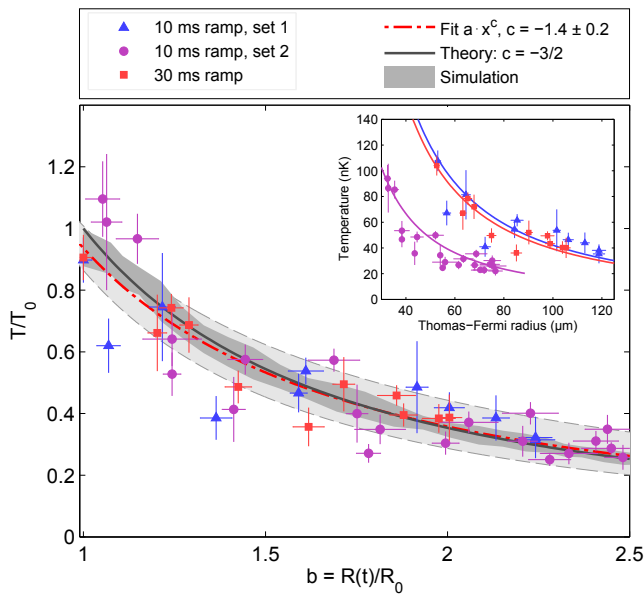} \caption{Temperature as a power law of the scaling factor. Main figure: datasets presented in figure \ref{fig2}(a) (blue triangles), \ref{fig2}(b) (purple circles) and \ref{fig2}(c) (red squares), respectively, recast in units of the initial temperature as a function of the scaling factor. Inset: data in absolute units. Vertical error bars are standard errors resulting from a bootstrapping method as applied in \cite{Kuhnert2013}. Horizontal error bars correspond to the error of measured cloud widths, normalised to the initial width. Dash-dotted line: power law fit to the data, with the lightly shaded area representing the fit's 95 \% confidence bounds. Black line: scaling model. Dark shaded area: classical field simulation with 120 sets of stochastic initial conditions generated by a SGPE. The plotted data is corrected for the independently measured heating rate.}
  \label{fig3}
 \end{figure}
 
The scale invariance of the underlying Hamiltonian allows to calculate time-dependent correlation functions: In the Thomas-Fermi regime, the density profile exhibits self-similar scaling described by
\begin{equation}
n(z,t) = \left(\frac{n_0}{b}\right)\left(1 - \frac{z^2}{R_0^2b^2}\right)\Theta\left(1 - \frac{|z|}{R_0b}\right),
\label{eqn_density}
\end{equation}
with a time-dependent scale factor $b(t) = R(t)/R_0$. Here, $R_0$ and $n_0$ denote the initial Thomas-Fermi radius and peak density, respectively, $\Theta$ is the Heaviside function and $z$ represents the axial coordinate. The scale factor obeys an Ermakov-like equation \cite{Chen2010}
\begin{equation}
\label{eqn:ermakov}
\ddot{b} + \omega_a^2(t)b = \frac{\omega_a(0)^2}{b^2}.
\end{equation}
Using the rescaled mean-field density (\ref{eqn_density}), we can write the linearised hydrodynamic equations for density and velocity fluctuations $\delta n$ and $\delta v$, disregarding the quantum pressure term, as
\begin{equation}
\frac{\partial}{\partial t}\delta n + \frac{\dot{b}}{b}\left(\delta n + z\frac{\partial}{\partial z}\delta n\right) = - \frac{n_0}{b}\frac{\partial}{\partial z}\left[\left(1 - \frac{z^2}{R_0^2b^2}\right)\delta v\right],
\end{equation}
and
\begin{equation}
\frac{\partial}{\partial t}\delta v + \frac{\dot{b}}{b}\left(\delta v + z\frac{\partial}{\partial z}\delta v \right)= - \frac{g}{m}\frac{\partial}{\partial z}\delta n.
\end{equation}
To solve these equations we introduce an ansatz of rescaled eigenmodes for density and phase fluctuations. This approach yields a set of uncoupled equations and hence no mixing of modes, finally predicting an adiabatic time evolution of the corresponding occupation numbers. For a thermal state, the initial phonon occupation numbers are given by a Bose distribution
\begin{equation}
\mathcal{N}_l(t = 0) = \frac{1}{\exp{\left[\frac{\hbar\omega_l\left(0\right)}{k_BT}\right]} - 1}.
\end{equation}
Adiabaticity results in a constant ratio $\omega_l(t)/T(t) = \omega_l(0)/T(0)$. The spectrum at $t = 0$ is given by \cite{Petrov2000}
\begin{equation}
\omega_l\left(0\right) = \frac{\omega_a}{\sqrt{2}}\sqrt{l\left(l+1\right)} = \frac{c_0}{R_0}\sqrt{l\left(l+1\right)},
\label{eqn_spectrum}
\end{equation}
with mode index $l$ and initial sound velocity $c_0$. For $t > 0$, it scales as $\omega_l(t) = \omega_l(0)b^{-3/2}$, due to the time-dependence of the sound velocity $c(t) = c_0/\sqrt{b}$ and radius $R(t) = R_0 b(t)$. Hence, for an initial state in thermal equilibrium, we obtain the temperature scaling
\begin{equation}
\label{eqn_tscaling}
T\left(t\right) = T\left(0\right)b^{-3/2}.
\end{equation}
The density correlations in free expansion that our thermometry scheme relies on are governed by the coherence function. For a thermal state with homogeneous density, as realised in the vicinity of the cloud center, it has the form \cite{Petrov2000, Reichel2010}:
\begin{equation}
\label{g1}
g^{\left(1\right)}(z,0) \simeq n\left(z,0\right)\exp{\left(-\frac{mk_BT|z|}{2 n(z,0)\hbar^2}\right)}, 
\end{equation}
where $n(z,0)$ denotes the density at time $t = 0$ and $k_B$ the Boltzmann constant. Based on our model, the coherence function is expected to scale as
\begin{equation}
\label{eqn_g1_sc}
\tilde{g}^{\left(1\right)}(z,t) \simeq \frac{n\left(z,0\right)}{b}\exp{\left[-\sqrt{\frac{1}{b}}\frac{mk_BT\left(0\right)|z|}{2{n\left(z,0\right)}\hbar^2} + \frac{im\dot{b}}{2\hbar b}z^2\right]}.
\end{equation}

 \begin{figure} 
   \includegraphics[width=0.48\textwidth]{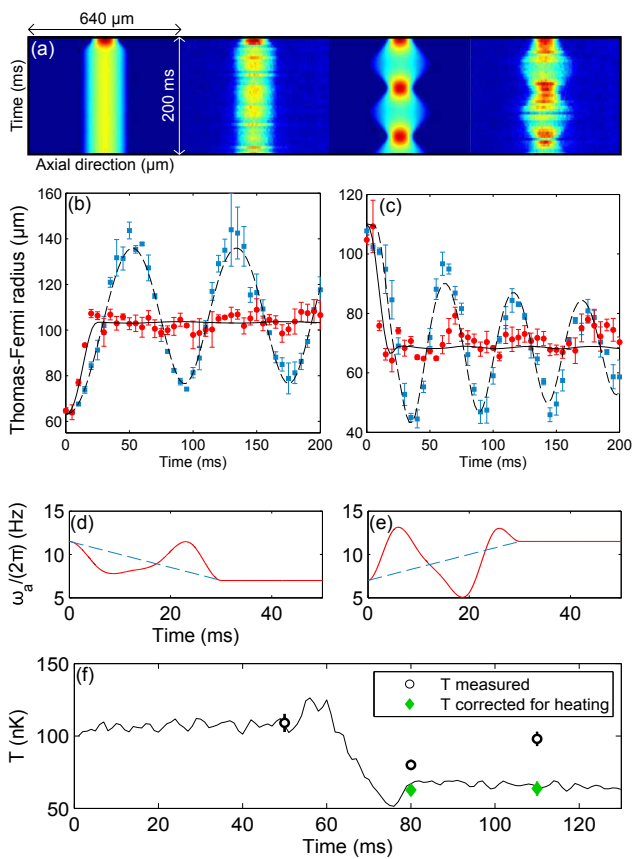}%
   \caption{STA for fast confinement changes. (a) Density profiles for optimal and linear ramps in simulation and experiment. Experimental profiles are averaged of 5 shots at identical parameters, taken at a free expansion time of 5 ms. (b) Measured Thomas-Fermi radii for an optimal decompression (red circles) from $\omega_a^0 = 2\pi\times 11.5$ Hz, $\omega_r^0 = 2\pi\times 764$ Hz to $\omega_a^f = 2\pi\times 7$ Hz, $\omega_r^f = 2\pi\times 1262$ Hz, and for a linear ramp (blue squares) compared with results from a GPE simulation (black and black-dashed lines). (c) Measured Thomas-Fermi radii after an optimal (red circles), and a linear ramp (blue squares) for a compression of the cloud, inverting initial and final trap frequencies as given for panel (b), again compared with GPE simulation results including damping (black and black-dashed lines). (d),(e) Optimal trap frequency ramp for decompression (d) and compression (e) within 30 ms (red line). Dashed lines: corresponding linear ramps. (f) Temperature measurements before and after the STA with (green diamonds) and without correction for extrinsic heating rate (black circles), compared to simulation results (black line).}
    \label{fig4}
   \end{figure}   

Figure \ref{fig3} summarizes the first central result of our experiments: The inset shows absolute temperatures plotted against measured Thomas-Fermi radii. If the measured temperatures are scaled to the initial temperature and plotted against the scale parameter $b(t) = R(t)/R_0$, the datasets collapse onto a single line.  This illustrates a scaling behaviour that is universal in sense that it is independent of absolute temperature, density or quench time. 
To validate our results we furthermore performed numerical simulations based on a stochastic Gross-Pitaevskii equation (SGPE) \cite{Stoof1999,Duine2001,Gardiner2002,Cockburn2011}, showing excellent agreement with the scaling model (fig. \ref{fig3}).

So far, we considered the dynamics induced by a linear ramp of the trapping potential. In the following, we demonstrate the conservation of phonon occupation numbers during shortcuts to adiabaticity \cite{Chen2010, DelCampo2011, Schaff2011a} for the rapid expansion and compression of a 1d quasi-BEC. 
To implement these shortcuts, we make use of an optimal control approach that is in spirit similar to the method proposed in \cite{Caneva2011}. We numerically solve the time-dependent 1d GPE with a suitable parametrisation of the trap which is subject to a global optimization procedure based on a genetic algorithm \cite{Holland1992, Rohringer2008}. The ramp speed is limited by the requirement of adiabaticity in the transverse degree of freedom. This constraint also guarantees that the gas remains in the 1d hydrodynamic regime, and that the interaction strength varies slowly with time. The properties of the ultracold gas therefore remain consistent with the conditions necessary for the validity of the microscopic scaling laws \cite{Gritsev2010} throughout the ramp.
 
The upper panel in figure \ref{fig4} shows a comparison between simulation and experiment for a linear and a shortcut ramp performing a decompression within 30 ms from a trap with frequencies $\omega_a^0 = 2\pi \times 11.5$ Hz and $\omega_r^0 = 2\pi \times 764$ Hz to $\omega_a^f = 2\pi \times 7$ Hz and $\omega_r^f = 2\pi \times 1262$ Hz. The subsequent dynamics is observed throughout a period of 170 ms, each picture taken after a short free expansion time of 5 ms, showing excellent agreement with simulations. It is interesting to note that our shortcut ramps are similar to theoretical results derived from a counter-diabatic driving method reported recently \cite{DelCampo2013}.  

For the STA, we expect an adiabatic state change, defined by $T/T_0 = \omega_a^f/\omega_a^0$. The temperature measurements, corrected for the measured heating rate, are in good agreement with the adiabatic prediction of $T/T_0 \approx 0.609$ for the implemented decompression shortcut, confirming that there is no additional heating during the applied procedure.

\section{Conclusion}

In summary, we have characterised the temperature of the phonon ensemble in a breathing quasi-1d Bose gas for different initial conditions, and used it to test the predicted dynamical scale invariance in the excitation spectrum of a quasi-1d Bose gas. Following these scaling laws, we have experimentally demonstrated rapid adiabatic expansion and compression of a 1d Bose gas in the hydrodynamic regime, allowing fast transformation of the trapped cloud without additional heating. 

Our work is only the beginning for studies of many-body scaling solutions and shortcuts to adiabaticity. The existence of scaling solutions has been proposed for a large class of cold atom systems \cite{Gritsev2010}. In principle, this opens up the interesting possibility to apply the techniques applied here to a variety of settings, such as fermionic systems or the 1d Bose gas with intermediate or strong interactions. We expect that studying the effect of quasiparticle interactions on the implementation of shortcuts to adiabaticity will shed new light on the complex many-body dynamics in these systems, in addition to providing novel tools for their controlled manipulation.

We expect that such extensions to studies in regimes of greater interaction strength, and to systems out of thermal equilibrium, will benefit from the tools presented in this work.
\section{Methods}

\paragraph{\textbf{Condensate preparation and detection.}} We employ standard cooling and magnetic trapping techniques \cite{Wildermuth2004} to prepare ultracold quasi-one-dimensional samples of \textsuperscript{87}Rubidium atoms in the $|F = 2, m_F = 2>$ state on an atom chip \cite{Reichel2010, Folman2000}. Atom chips feature microfabricated wire structures to create fields for atom trapping and manipulation \cite{Groth2004}. The structures used in our experiments are produced by masked vapor depositon of a 2 $\mu m$ gold layer on a silicon substrate, with a width of both trapping and control wires of 200 $\mu m$. For detection, we employ resonant absorption imaging \cite{Smith2011} using a high quantum-efficiency CCD camera (Andor iKon-M 934 BR-DD) and a diffraction-limited optical imaging system characterised by an Airy radius of $~4.5$ $\mu m$. The RF shield at 12 kHz above the bottom of the trap is used to limit the number of atoms in the thermal background cloud populating transverse excited states of the trap, which would otherwise adversely affect our thermometry scheme by reduction of interference contrast in free expansion.\\

\paragraph{\textbf{Characterization of the breathing mode.}}

\begin{figure} %\hspace*{-10px}
	\includegraphics[width=0.48\textwidth]{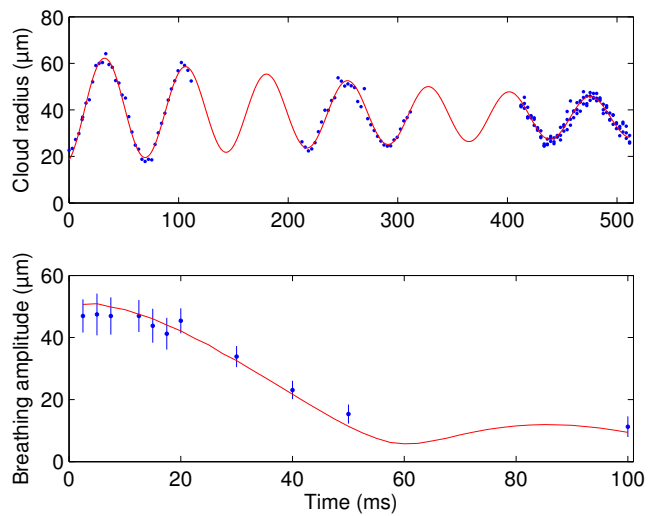}%
	\caption{Characterization of the breathing mode induced by a trap quench. Upper panel: Breathing induced by a linear quench during a time $\tau =$ 12.5 ms. The fit includes an exponential damping term, with a time constant $1/\lambda =$ 500 ms. Lower panel: Breathing amplitude plotted against quench time $\tau$. Error bars correspond to 95 \% confidence intervals of fits as shown in the upper panel. The theoretical calculations (line) are based on numerically solving a 1d GPE.}
	\label{fig5}
\end{figure}

We characterise the breathing mode excited by a linear trap frequency ramp from $\omega_a =$ $2\pi\times 12.1$ Hz to $2\pi\times 8.2$ Hz, and $\omega_r = $ $2\pi\times 630$ Hz to $2\pi\times 990$ Hz, respectively, in figure \ref{fig5}. As an example, the upper panel shows the time evolution of the cloud radius after a ramp with duration $\tau =$ 12.5 ms. Fitting data as presented here allows us to extract frequencies, damping rates and amplitudes of the breathing mode. The frequency $\omega_b$ is influenced by the total atom number in the trap, and is expected to vary with the axial trap frequency between $\omega_b/\omega_a = \sqrt{3}$ in the 1d limit, and  $\omega_b/\omega_a = \sqrt{2.5}$ representing the elongated 3d regime \cite{Menotti2002}. The amplitude strongly depends on the duration and shape of the trap frequency ramp. The lower panel in figure \ref{fig5} shows a comparison of measured breathing amplitudes for different ramp times between 2 ms and 100 ms with results calculated with a 1d Gross-Pitaevskii equation (GPE), taking into account corrections to the interaction term relevant in the 1d / 3d crossover regime \cite{Gerbier2004}, and shows good agreement in the chosen parameter range.  

\paragraph{\textbf{Thermometry.}} In this work we use the thermometry scheme proposed and demonstrated in \cite{Imambekov2009, Manz2010} based on the analysis of density correlations in freely expanding phase-fluctuating quasi-1d condensates and comparison with numerically calculated density profiles \cite{Stimming2010}. 

Breathing contributes a velocity field characterized by the derivative of the scale parameter $\dot{b}$, leading to an additional axial compression or expansion of the density profile during free expansion. This effect can be accounted for by an additional phase factor
\begin{equation}
\Psi(z) = \psi(z)\exp{\left[\frac{im}{4\hbar}\frac{\dot{b}}{b} z^2\right]}
\end{equation}
in the numerics, where $b$ and $\dot{b}$ are determined by fits to the measured breathing oscillations. The error on the temperature measurements is estimated by a bootstrapping method as outlined in \cite{Kuhnert2013}. \\

\paragraph{\textbf{Derivation of the temperature scaling.}}
The general conditions for the existence of a scaling solution are stated in reference \cite{Gritsev2010}. For the 1d Bose gas, they are fulfilled in the presence of contact interactions, as well as a harmonic, linear or vanishing axial trapping potential. Given that our system is a 1d quasicondensate, and the trapping potential is harmonic, we can derive the corresponding hydrodynamic scaling relations for correlation functions.
Our starting point is the self-similar scaling of the density profile:
\begin{equation}
n(z,t) = \frac{n_0}{b}\left(1 - \frac{z^2}{R_0^2b^2}\right)\Theta\left(1 - \frac{|z|}{R_0b}\right).
\label{eqn_density}
\end{equation}
$\Theta$ denotes the Heaviside function, $R_0$ the initial Thomas-Fermi radius and $b$ the scale parameter. Similar to the discussion of the corresponding equilibrium problem \cite{Petrov2000}, a scaling solution in terms of eigenmodes for density and velocity fluctuations $\delta n$ and $\delta v$ can be formulated as
\begin{equation}
\delta n = \frac{1}{b}\sum\limits_{l=1}^{\infty}{P_l\left(\tilde{z}\right)A_l\cos{\eta_l}}
\end{equation}
and
\begin{equation}
\delta v = -\sqrt{\frac{1}{b}}\sum\limits_{l=1}^{\infty}{\frac{g}{mR_0\omega_l(0)}\frac{d}{d\tilde{z}}P_l\left(\tilde{z}\right)A_l\sin{\eta_l}},
\end{equation}
with the Legendre polynomials $P_l(\tilde{z})$, the interaction constant $g$, rescaled coordinates $\tilde{z} = z/R = z/(R_0b)$, and time-dependent amplitudes $A_l\sin{\eta_l}$ and $A_l\cos{\eta_l}$. $\eta_l$ denotes the frequency of the oscillation between the quadratures of the mode $l$. Correspondingly, the initial equilibrium spectrum scales as 
\begin{displaymath}
\omega_l(t) = \omega_l(0)b^{-3/2}.
\end{displaymath}
Substituting $\delta n$ and $\delta v$ into the linearised Euler equations
\begin{equation}
\frac{\partial}{\partial t}\delta n + \frac{\dot{b}}{b}\delta n + \frac{\dot{b}}{b}z\frac{\partial}{\partial z}\delta n = - \frac{n_0}{b}\frac{\partial}{\partial z}\left[\left(1 - \frac{z^2}{R_0^2b^2}\right)\delta v\right]
\end{equation}
\begin{equation}
\frac{\partial}{\partial t}\delta v + \frac{\dot{b}}{b}\delta v + \frac{\dot{b}}{b}z\frac{\partial}{\partial z}\delta v = - \frac{g}{m}\frac{\partial}{\partial z}\delta n,
\end{equation}
where we have disregarded the quantum pressure term, yields
\begin{equation}
\dot{\eta}_l = \omega_l\left(t\right) - \frac{1}{2}\frac{\dot{b}}{b}\sin{\eta_l}\cos{\eta_l},\;\;\frac{\dot{A}_l}{A_l} = -\frac{1}{2}\frac{\dot{b}}{b}\sin^2{\eta_l}.
\end{equation}
Since the characteristic inverse time scale of the breathing mode $(\dot{b}/b)_{max} = \omega_a/2$ is small compared to the characteristic frequencies $\omega_l\left(0\right)$ of the phonon modes with $l > 2$, we can average over rapid oscillations of $\cos \eta_{l}$ and $\sin \eta_{l}$ to reduce these expressions to $\dot{\eta}_l \simeq \omega_l\left(t\right)$ and $A_l \simeq A_l\left(0\right)(R_0/R)^{1/4}$, and the phonon modes are expected to scale adiabatically. Then the initial number of phonons in a thermal state,
\begin{equation}
\mathcal{N}_l(t) = \frac{1}{\exp{\frac{\hbar\omega_l\left(t\right)}{k_BT\left(t\right)}} - 1} = \mathcal{N}_l(0)
\end{equation}
is conserved, resulting in
\begin{equation}
\frac{\omega_l\left(t\right)}{T\left(t\right)} = \frac{\omega_l\left(0\right)}{T\left(0\right)}.
\end{equation}
This leads to the observed temperature scaling $T(t) = T(0)b^{-3/2}$.

The decay of the coherence function of a quasicondensate is dominated by phase noise \cite{Mora2003}. We can express phase fluctuations in terms of velocity fluctuations using the relation
\begin{displaymath}
\delta\phi_l\left(z,t\right) = R\sqrt{\frac{1}{b}}\frac{m}{\hbar}\int_{0}^{\tilde{z}}{d\tilde{z}'\delta\tilde{v}\left(\tilde{z}'\right)A_l\sin{\eta_l}},
\end{displaymath}
where
\begin{displaymath}
\delta\tilde{v} = \frac{g}{mR_0\omega_l(0)}\frac{d}{d\tilde{z}}P_l\left(\tilde{z}\right),\;\; A_l \simeq A_l\left(0\right)(R_0/R)^{1/4}. 
\end{displaymath}
Therefore the relation between initial and time-dependent modes $\delta\phi_l$ reads
\begin{equation}
\label{eqn_phasefluct}
\delta\phi_l\left(z,t\right) = b^{1/4}\delta\phi_l\left(z/b,0\right)\frac{\sin{\eta_l\left(t\right)}}{\sin{\eta_l\left(0\right)}}.
\end{equation}
The time-dependent one-body reduced density matrix can be expressed as
\begin{equation}
\label{eqn_obdm}
\rho\left(z,z',t\right) = \sqrt{nn'}\exp{\left[-\frac{1}{2}\left\langle\delta\phi^2_{zz'}\right\rangle + \frac{im\dot{b}}{2\hbar b}\left(z^2-z'^2\right)\right]},
\end{equation}
with
\begin{displaymath}
\left\langle\delta\phi^2_{zz'}\right\rangle = \left\langle\left[\delta\phi\left(z,t\right) - \delta\phi\left(z',t\right)\right]^2\right\rangle,
\end{displaymath}
as well as $n = n(z)$ and $n' = n(z')$. Using $\delta\phi = \sum_{l}\phi_l$, we can write the density matrix (\ref{eqn_obdm}) in terms of the modes given in equation (\ref{eqn_phasefluct}). Substituting and following the steps in reference \cite{Mora2003}, we find that near the cloud center, where the density is practically uniform and we can use trigonometric approximations for $P_l$ \cite{Abramowitz1964},
\begin{equation}
\rho\left(z,z',t\right) \simeq \frac{\sqrt{nn'}}{b}\exp{\left[-\frac{|z-z'|}{\sqrt{b}\lambda_T}+\frac{im\dot{b}}{2\hbar b}\left(z^2-z'^2\right)\right]},
\end{equation}
with a coherence length $\lambda_T = \frac{2{n\left(z\right)}\hbar^2}{mk_BT(0)}$. This corresponds to a transformation of the form
\begin{displaymath}
\rho\left(z,z',t\right) = \frac{1}{b}\rho\left(\frac{z}{\sqrt{b}},\frac{z'}{\sqrt{b}},0\right)\exp\left[-iF(t)\left(z^2-z'^2\right)\right],
\end{displaymath}
as predicted in reference \cite{Gritsev2010}, with the difference that the spatial coordinates scale with $b^{-1/2}$ instead of $b^{-1}$. This difference is a consequence of the the Thomas-Fermi approximation. In the hydrodynamic regime, scale invariance therefore holds even if the interaction strength is kept constant. In contrast, reference \cite{Gritsev2010} assumes a suitable tuning of the interaction constant, thereby yielding an exact solution valid for arbitrary values of the Lieb-Liniger parameter.  \\

\paragraph{\textbf{Heating.}} The temperature scaling $T = T_0b^{-3/2}$ satisfies the equation
\begin{equation}
\frac{\dot{T}}{T} = -\frac{3}{2}\frac{\dot{b}}{b}.
\end{equation}
In our experiment we observe heating during evolution times of several hundreds of milliseconds. We find that all our measurements are compatible with a linear increase of temperature over time, which can be represented by adding a constant heating term to the equation:
\begin{equation}
\dot{T} = -\frac{3}{2}\frac{\dot{b}}{b}T + \alpha T_0.
\end{equation}
This equation is solved by $T = T_0h(t)b^{-3/2}$, with $h(t)$ given by
\begin{equation}
h(t) = 1 + \alpha\int_{0}^{t}dt'b\left(t'\right)^{3/2}.
\end{equation}
The integral can be calculated numerically and $\alpha$ corresponds to the regular heating rate in units of the initial temperature for constant $b$. \\

\paragraph{\textbf{Finite temperature simulations.}} We solve a stochastic 1d Gross-Pitaevskii equation (SGPE) \cite{Stoof1999,Duine2001,Gardiner2002,Cockburn2011} of the form
\begin{equation}
i\hbar\frac{\partial\psi}{\partial t} = \left[1 - i\gamma\left(T\right)\right]\left[H_{GP} - \mu\right]\psi + \eta,
\end{equation}
where
\begin{equation}
H_{GP} = -\frac{\hbar^2}{2m}\frac{\partial^2}{\partial z^2} + \frac{1}{2}m\omega_a^2z^2 + g_{1d}|\psi|^2.
\end{equation}
Here $\mu$ denotes an external chemical potential, and $\gamma(T)$ is a damping coefficient that is coupled to the $\delta$-correlated noise term $\eta$ via a fluctuation-dissipation theorem:
\begin{equation}
\langle\eta^{*}\left(z,t\right)\eta\left(z', t'\right)\rangle = 2\hbar k_BT\gamma\left(T\right)\delta\left(t - t'\right)\delta\left(z - z'\right).
\end{equation}

Repeated solution of the SGPE yields a set of independent wave functions representing a thermal state. We use this state as initial condition for propagation with a time-dependent Gross-Pitaevskii Hamiltonian without any noise or damping terms. Such an approach has previously been applied to model condensate formation in atom chip traps \cite{Proukakis2006} and is very similar to other classical field methods based on stochastic sampling of initial conditions \cite{Sinatra2002, Witkowska2010}. The simulation results are analysed with the same procedures as the experimental data.

\section{Acknowledgements}
We are grateful to M. Wilzbach, D. Heine and B. Hessmo for initial work building the experimental apparatus. We thank J-F. Schaff, N. Proukakis, P. Gri\v{s}ins and B. Rauer for fruitful discussions.
This work was supported by the Austrian FWF through the Wittgenstein Prize, the Doctoral Programme CoQuS (W1210), the SFB FoQuS (F4010-N23), the FFG project PLATON, and the EU through the projects QuantumRelax (ERC-ADG-320975) and SIQS.
I.E.M. acknowledges the financial support from the FWF (project P22590-N16).

\section{Author contributions}
W.R. and D.F. performed the experiments and analysed the data. F.S. contributed to building the experiment and provided help with the data analysis. I.E.M. provided important advice for the execution of the scaling measurements and developed the theoretical model. W.R. devised the optimal control scheme and performed numerical simulations. J. S. and M. T. provided scientific guidance and funding for the experiment. All authors contributed to the interpretation of the data and the writing of the manuscript.

\section{Additional information}
The authors declare no competing financial interests.

\end{document}